\documentclass[superscriptaddress,aps,pra,twocolumn,showpacs,nofootinbib,longbibliography,notitlepage]{revtex4-2}
\usepackage{etex}
\usepackage{amsmath,amssymb,amsthm}
\usepackage[colorlinks=true,citecolor=blue,urlcolor=blue]{hyperref}
\usepackage[pdftex]{graphicx}
\usepackage{times,txfonts}
\usepackage{braket}
\usepackage{color}
\usepackage{natbib}
\usepackage{amsmath,blkarray}
\usepackage{mathtools}
\usepackage{latexsym}
\usepackage{tabularx, booktabs}
\usepackage{graphics,epstopdf}
\usepackage{graphicx}
\usepackage{float}
\usepackage{graphicx}
\usepackage{amsfonts}
\usepackage{subcaption}
\usepackage{color,soul}

\newcommand{\be}{\begin{equation}}
	\newcommand{\ee}{\end{equation}}
\newcommand{\ba}{\begin{eqnarray}}
	\newcommand{\ea}{\end{eqnarray}}

\begin{document}

	\title{Quantum violation of trivial and non-trivial preparation non-contextuality: Nonlocality and Steering}
	\author{ Prabuddha Roy }
	\author{ A. K. Pan }
	\email{prabuddharoy.94@gmail.com}
	\email{akp@phy.iith.ac.in}
	\affiliation{National Institute of Technology Patna, Ashok Rajpath, Patna, Bihar 800005, India}
	\affiliation{Department of Physics, Indian Institute of Technology Hyderabad, Telengana-502284, India }

\begin{abstract}
This paper illustrates a direct connection between quantum steering and non-trivial preparation contextuality. In two party-two measurement per party-two outcomes per measurement $(2-2-2)$ Bell scenario, any argument of Bell nonlocality is a proof of trivial preparation contextuality; however, the converse may not hold. If one of the parties (say, Alice) performs the measurements of more than two dichotomic observables, then it is possible to find a set of non-trivial functional relations between Alice's observables. We argue that the existence of a suitable set of such non-trivial relations between Alice's observables may warrant the unsteerability of quantum states at the end of another spatially separated party (say, Bob). Interestingly, such constraints can be read as non-trivial preparation non-contextuality assumptions in an ontological model. We further demonstrate two types of Bell inequalities that can be converted into linear steering inequalities using the aforementioned non-trivial conditions on Alice's observables. Such steering inequalities can also be considered as non-trivial preparation noncontextual
inequalities. Since the local bound of the family of Bell expression gets reduced under the additional non-trivial conditions, it provides a test of quantum steering and nonlocality from the same family of Bell expressions depending upon its violation of the non-trivial preparation non-contextual or the local bound, thereby establishing a direct connection between quantum steering and non-trivial preparation contextuality.
\end{abstract}
\maketitle
\section{introduction}
No-go theorems play a significant role in quantum foundations research by providing a route to examine the constraints required to be satisfied by an ontological model to be consistent with the statistical prediction of quantum theory. Any ontological model seeks to provide a `complete description' of a quantum system's state so that an observable's individual measured values are predicted by an appropriate set of ontic states (usually denoted as $\lambda$'s). Studies along this line have resulted in important discoveries. Bell's theorem \cite{bell64,bell66} is the first that demonstrates a conflict between the quantum theory and an ontological model satisfying locality. The Kochen and Specker (KS) theorem \cite{ks} proves an inconsistency between the quantum theory and the non-contextual ontological model. 

Quantum nonlocality found several \cite{marco05,Brunner2014,marco19} application in device-independent cryptography \cite{ekert, acin,bar13,zap22}, certification of randomness \cite{pir10,acin12,col12,pathak22}, self-testing \cite{supic}, certification of dimension of Hilbert space \cite{dim,mahato}. On the other hand, quantum contextuality also found applications in providing an advantage in communication games \cite{spek09,hameedi,ghorai}, in certification and generation of randomness \cite{pan21d,um13,um20,pathak22} and in quantum computation \cite{how,rau}. While the demonstration of Bell's theorem requires two or more space-like separated systems, the KS theorem requires a single system with the Hilbert space dimension $d\geq 3$ \cite{cabello,ker,yu,mermin,cabello08,pan10,kly,pan21d,pan21s,budroni21}. The original KS proof was demonstrated by using 117 projectors for the qutrit system. Later, simpler versions and variants of it using a lower number of projectors have been provided \cite{cabello08,pan10}. 

The traditional KS proof of contextuality has a limited scope of applicability for the following reasons. First, along with measurement non-contextuality, KS proof requires outcome determinism for sharp measurement in an ontological model. Second, it is not the model of arbitrary operational theory, somewhat specific to quantum theory. Third, it does not apply to the generalized measurements, i.e., for POVMs. In this regard, importantly, it is to be noted that the traditional notion of KS non-contextuality has been generalized by Spekkens \cite{spek05,hari,spek14} for any arbitrary operational theory and extended the formulation to the transformation and preparation of non-contextuality. 

An ontological model is preparation non-contextual if two operationally equivalent preparation procedures are represented by the same ontic state in that model \cite{spek05,pan21s,kumari}. In Spekken's proof of preparation and measurement quantum contextuality for qubit system, a set of trivial (originates by construction)  along with a set of non-trivial (dictates by the suitable choices of measurement observables) non-contextuality assumptions are required.  

Another form of quantum correlation known as quantum steering has recently attracted many attention \cite{epr,sch,reid,wise1,jones,cava, pati,uolarev,quin,uolaprl,dd1,dd2}. Given a suitable entangled state,  quantum steering can be revealed between the outcomes of Alice's measurements and the resulting post-measured states of Bob. It is a weaker form of quantum correlation than nonlocality but stronger than quantum entanglement.

For the 2-2-2 Bell scenario the violation of locality can be shown to be equivalent to \emph{trivial} preparation contextuality \cite{barrett,pusey,tava19}.  This work aims to establish a connection between quantum steering and the quantum violation of \emph{non-trivial} preparation non-contextuality. Given a suitable entangled state, if  Alice performs the measurements of more than two dichotomic observables, then there is the possibility of finding a suitable set of  \emph{non-trivial} functional relations between Alice's observables. The existence of a suitable set of such non-trivial relations between Alice's observables may warrant the unsteerability of quantum states at Bob's end. In an ontological model, such constraints on Alice's observables provide a set of non-trivial preparation non-contextuality assumptions. Further, by considering two types of Bell inequality, we have demonstrated that the local bounds of those inequalities get reduced and become non-trivial preparation non-contextuality inequalities if non-trivial conditions on Alice's observables are imposed. Thus, the non-trivial preparation non-contextuality inequalities derived here can be considered as steering inequalities. It is then simple to understand that given an entangled state and incompatible measurements, the quantum value of the Bell inequalities may not reveal nonlocality but may demonstrate a form of quantum inseparability through quantum steering or preparation contextuality.

In order to demonstrate our results, we first consider a Bell's inequality in a bipartite scenario involving three dichotomic observables for each party and demonstrate how it can be converted into a steering (or non-trivial preparation non-contextual) inequality by using the non-trivial functional relation between Alice's observables. Further, we consider the well-known Gisin's elegant Bell inequality \cite{gisin} where Alice and Bob measure four and three dichotomic observables, respectively. Using it, we demonstrate how this local Bell inequality can be converted into a known linear steering inequality. Hence, the steering inequalities derived here can be considered as non-trivial preparation non-contextuality inequalities. Our work thus tests quantum steering and nonlocality from the same family of Bell expressions depending on its quantum violations.

 \section{preliminaries}

Let us encapsulate the notions of quantum steering and the preparation non-contextuality in an ontological model of quantum theory.

\subsection{Ontological model of quantum theory and preparation non-contextuality}
Harrigan and Spekkens \cite{spek05, hari} coherently formulated the ontological model of an operational theory. Let there is a set of preparation procedures $\mathcal{P}$, a set of measurement procedures $\mathcal{M}$ and a set of outcomes $\mathcal{K}_{M}$. Given a preparation procedure $P\in \mathcal{P}$  and a measurement procedures $M\in \mathcal{M}$, an operational theory assigns probability $p(k|P, M)$ of occurrence of a particular outcome $k\in \mathcal{K}_{M}$. For example, in quantum theory, a preparation procedure produces a density matrix $\rho$, and the measurement procedure is generally described by a suitable POVM $E_k$. The probability of occurrence of a particular outcome $ k $ is determined by the Born rule, i.e., $p(k|P, M)=Tr[\rho E_{k}]$. Here we restrict our discussion to a particular operational theory, i.e., quantum theory. 

In an ontological model of the quantum theory, it is assumed that whenever $\rho$ is prepared by a preparation procedure $P\in \mathcal{P}$ a probability distribution $\mu_{P}(\lambda|\rho)$ in the ontic space is prepared, satisfying $\int _\Lambda \mu_{P}(\lambda|\rho)d\lambda=1$ where $\lambda \in \Lambda$ and $\Lambda$ is the ontic state space. The outcome $k$ is distributed as a response function $\xi_{M}(k|\lambda, E_{k}) $ satisfying $\sum_{k}\xi_{M}(k|\lambda, E_{k})=1$ where a PO,VM $E_{k}$ is realized through a measurement procedure $M\in\mathcal{M}$. A viable ontological model of quantum theory should reproduce the Born rule, i.e., 
\begin{align}
	\forall \rho , \forall E_{k}, \forall k, \ \ \int _\Lambda \mu_{P}(\lambda|\rho) \xi_{M}(k|\lambda, E_{k}) d\lambda =Tr[\rho E_{k}]
\end{align}

An ontological model of quantum theory is assumed to be measurement non-contextual if 

\begin{align}
	\forall P, \ \  p(k|P, M)=p(k|P, M^{\prime})\\
	\nonumber
	\Rightarrow   \ \ \xi_{M}(k|\lambda, E_{k})=\xi_{M^{\prime}}(k|\lambda, E_{k})
\end{align}
where $M$ and $M^{\prime}$ are two measurement procedures realizing the same POVM, $E_k$. Note that KS  non-contextuality assumes the measurement non-contextuality for the sharp measurements along with the deterministic response functions for projectors \cite{spek05}. An ontological model of quantum theory can be considered to be preparation non-contextual if  

\begin{align}
	\forall M  , \ \  &p(k|P, M)=p(k|P^{\prime}, M)\\
	\nonumber
	\Rightarrow &\mu_{P}(\lambda|\rho)=\mu_{P^{\prime}}(\lambda|\rho)
\end{align}
 where $P$ and $P^{\prime}$ are two distinct preparation procedures preparing the same density matrix $\rho$. We refer to Spekken's paper \cite{spek05} for more details regarding this issue.

For our purpose, we distinguish the assumptions preparation non-contextuality in an ontological model of quantum theory as trivial and non-trivial ones \cite{ghorai}. The non-trivial conditions only arise if the density matrix $\rho$ is prepared through the measurements of three or more observables. Although we have provided detailed examples later, a small clarification could be helpful for the reader. 

Let Alice and Bob share a maximally entangled state. Each of Alice's projective measurements will produce a maximally mixed state $\mathbb{I}/2$ at Bob's end. Preparation non-contextuality assumption dictates us to consider same distribution $\mu_{P_{x}}(\lambda|\mathbb{I}/2)$ for all preparation procedures $\{P_{x}\}\in \mathcal{P}$ where $x= \{1,2,3,. . . . , m\}$. If $\mathbb{I}/2$ is prepared through Alice's projective measurement, in this paper, we call them trivial preparation procedures leading to the trivial preparation non-contextuality assumptions in an ontological model. However, Alice can also prepare the density matrix $\mathbb{I}/2$ through the measurements of POVMs, which provides non-trivial preparation non-contextuality assumptions considered in this paper. We show that a set of non-trivial relations between Alice's observables can be found in such a case. Such non-trivial conditions enable one to provide logical proof of preparation contextuality \cite{spek05}. Note here that such non-trivial conditions satisfied by Alice's observables are the key to our argument as it provides the unsteerability of quantum states.

\subsection{Quantum steering}
Quantum steering is a different demonstration of action-at-a-distance than nonlocality. This notion is first introduced by Schr$\ddot{o}$dinger to capture the essence of the Einstein-Podolsky-Rosen argument. A few decades later, an experimentally testable criterion for EPR steering was formulated by Reid \cite{reid} using the Heisenberg uncertainty relation. Later, Wiseman and his co-workers \cite{wise1,jones,cava} demonstrated the concept of steering as a communication game and proved that the steering is incompatible with a local hidden state model. Let Alice and Bob be spatially separated observers who share a bipartite quantum state $\rho_{AB}$ and  Alice's (Bob's) measurement setting $x$ ($y$) produce output $a$ ($b$). Alice measures the POVMs $\{M_{a|x}\}$ satisfying $M_{a|x} \geq 0$ and $\sum_{a} M_{a|x} =\mathbb{I}$, and Bob measures POVMs $\{M_{b|y}\}$ satisfying $M_{b|y} \geq 0$ and $\sum_{b} M_{b|y} =\mathbb{I}$. In quantum theory, the joint probability distribution is given by $p(a b|x y) = Tr[\rho_{AB} M_{a|x}\otimes M_{b|y}]$. 

Now, in a local model $p(a b|x y)$ admits a decomposition of the following form
\begin{align}
\label{loc}
	p(a b|x y)=\int d\lambda \pi(\lambda) p_{A}(a|x, \lambda)p_{B}(b|y, \lambda)
\end{align}
where $\lambda$ is the hidden variable distributed as  $\pi(\lambda)$ and $p_{A}(a|x, \lambda)$ and $p_{B}(b|y, \lambda)$ are respective local probabilities of Alice and Bob.  A quantum state $\rho_{AB}$ is said to be non-local when the statistics of arbitrary local measurements do not admit a decomposition of the form of Eq. (\ref{loc}). In such a case, the correlations $	p(a b|x y)$ can be used to demonstrate the violation of a suitable Bell inequality for suitably chosen local measurements.

 In a steering scenario, Bob wants to verify if $\rho_{AB}$ is entangled. He asks Alice, whom he does not trust,  to perform measurement $x$ on her subsystem and declare her output $a$. After Alice's measurement, the un-normalized state of Bob's subsystem is $ 	\sigma_{a|x} = Tr_{A}[\rho_{AB} (M_{a|x} \otimes \mathbb{I})$. The corresponding normalized state is $\rho_{a|x}=\sigma_{a|x}/Tr[\sigma_{a|x}]$. We also have, $\sum_{a}\sigma_{a|x}=\sum_{a}\sigma_{a|x^{\prime}}$ when $x=x^{\prime}$ which ensure no signaling from Alice to Bob. Now, Bob has to ensure that the assemblage $\{\sigma_{a|x}\}$ does not admit a decomposition of the form

\begin{align}
\label{steer}
		\sigma_{a|x}= \int d\lambda \pi (\lambda) p_{A}(a|x, \lambda)\sigma_{\lambda},
\end{align}
which satisfy $\sum_{a}\sigma_{a|x}=\rho_{B} $.  If the assemblage $\{\sigma_{a|x}\}$ does not admit a decomposition of the form of Eq. (\ref{steer}), the state $\rho_{AB}$ is called steerable. In such a case, suitable steering inequalities can be violated. Such suitable steering inequalities are the well-known linear steering inequalities \cite{saunders} given as follows
 \begin{align}
\label{genst}
\frac{1}{n}\sum_{y=1}^{n} \langle A_{y}\otimes B_{y}\rangle \leq C_{n}
\end{align}
where $C_{n}=\max\limits_{\{A_{y}\}}\Bigg[\lambda_{max}\Big(\dfrac{1}{n}\sum_{y=1}^{n} \langle A_{y}\otimes B_{y}\rangle\Big)\Bigg]$ with $\lambda_{max}(\hat{O})$ denotes the largest eigenvalue of $O$. The violation of the inequality in Eq. (\ref{genst}) confirms that Alice can steer Bob's state. 

It is important to note here that the assemblage $\{\sigma_{a|x}\}$ is un-steerable for any state $\rho_{AB}$ if and only if the set of Alice's POVMs $\{M_{a|x}\}$ is jointly measurable \cite{quin,uolaprl}. The set $\{M_{a|x}\}$ is said to be jointly measurable iff there exists a grand POVM $\mathbb{G}=\{G(\lambda), 0\leq G(\lambda)\leq 1, \sum_{\lambda} G(\lambda)=\mathbb{I}\}$ for which $M_{a|x}$ can be obtained via classical post-processing. So that, $M_{a|x}=\sum_{\lambda} D_{\lambda} (a|x)G(\lambda)  \ \ \ \forall a,x$ where $D_{\lambda}(a|x)$  is a positive number and $\sum_{\lambda}D_{\lambda}(a|x)=1$. 

In the following section, we show that a family of suitable local Bell inequalities reduces to linear steering inequalities under the assumption of non-trivial preparation non-contextuality.

\section{Connection between steering and non-trivial quantum preparation contextuality}
\label{pncsteer}
Let Alice measures a set of dichotomic observables $\{A_{n,x}\}$  where $x= \{1,2,3,. . . . , n\}$, and corresponding projectors are $P_{A_{n,x}}^{a}=|x_{a}\rangle\langle x_{a}|=(\mathbb{I}+a A_{n,x})/2$ where $a\in \{+,-\}$ and $\mathbb{I}=P_{A_{n,x}}^{+} + P_{A_{n,x}}^{-}$. If Alice and Bob share a maximally entangled state  $\rho_{AB}\in \mathbb{C}^{2}\otimes \mathbb{C}^{2}$ then each of Alice's measurements produces a maximally mixed state to Bob's end is given by 
\begin{align}
\label{kk}
	\frac{\mathbb{I}}{2}=\frac{1}{2}P_{A_{n,x}}^{+} + \frac{1}{2} P_{A_{n,x}}^{-}
\end{align}
An equivalent representation of Eq. (\ref{kk}) in ontological model provides the assumption of \emph{trivial} preparation non-contextuality. For a two-party binary measurement Bell scenario,  the assumption of trivial preparation non-contextuality provides the locality assumption \cite{barrett,pusey}. 

Now, consider that the suitable choice of  Alice's observables also follow a non-trivial relation between them so that the equal mixture of all the positive or negative eigenvalue projectors of $A_{n,x}$ follow 
\begin{eqnarray}
\label{m}
	\left\{\frac{\mathbb{I}}{2},\frac{\mathbb{I}}{2}\right\}=\left\{\frac{1}{n} \sum_{x=1}^{n} P_{A_{n,x}}^{+} , \frac{1}{n} \sum_{x=1}^{n} P_{A_{n,x}}^{-} \right\}
\end{eqnarray}
which is valid only if a functional relationship $\sum_{x=1}^{n} A_{n,x}=0$ between the Alice's observables exists with $n\geq 3$. One can also think that the maximally mixed states in Eq. (\ref{m}) are produced if Alice's measurement is implemented by the measurements of suitable POVMs. Importantly, Eq. (\ref{m}) will also have an equivalent representation in an ontological model which we term as \emph{non-trivial} preparation-noncontextuality assumption originating from the functional relation between the observables. Note that the Eq. (\ref{m}) may not be unique. Any such case will provide a different set of non-trivial preparation non-contextuality assumptions in the ontological model of quantum theory.

 Measurements of Alice's observables produce  Bob's assemblage $\{\sigma_{a|x}=|x_{a}\rangle\langle x_{a}|/2n\}$ if the measurement is implemented by the way described in Eq. (\ref{m}). If the assemblage admits the decomposition given by  Eq. (\ref{steer}) then each of $\{|x_{a}\rangle\langle x_{a}|/2n\}$  must satisfy Eq. (\ref{steer}) \cite{pati,uolarev}. It can be seen that if Alice's observables satisfy Eq. (\ref{m}) then $\pi(\lambda)=1/2n$ for every $n\geq 3$, so that, $\sum_{\lambda}\pi(\lambda)=1$ . This implies that the non-trivial measurement procedure described in Eq. (\ref{m}) cannot steer Bob's assemblage. Interestingly, in an ontological model of quantum theory, the equivalent representation of Eq. (\ref{m})  is just the non-trivial preparation-noncontextuality assumptions. This hints that if a preparation non-contextuality inequality is derived by assuming the non-trivial conditions on Alice's observables, such inequality can be considered as a steering inequality.

The critical question is whether such choices of observables are available in quantum theory. In the following, we show this by providing two simple examples for the qubit system.

\section{Converting local Bell inequalities to steering inequalities}

\subsection{Example of trine-spin axes observables of Alice}
\label{sstrine}
Let us again consider that Alice and Bob share a maximally entangled state $\rho_{AB}\in \mathbb{C}^{2}\otimes \mathbb{C}^{2}$. Alice's preparation procedures $\{P_{x}\}\in \mathcal{P}$ to prepare the maximally mixed state $\mathbb{I}/{2}$ are implemented through the measurements of dichotomic observables $\{A_{3,x}\}$  where $x= \{1,2,3\}$. For our purpose, we consider the observables along trine-spin axes, so that,  $A_{3,1}=\sigma_{z}$, $A_{3,2}=\frac{\sqrt{3}}{2}\sigma_{x}-\frac{1}{2}\sigma_{z}$ and $A_{3,3}=\frac{-\sqrt{3}}{2}\sigma_{x}-\frac{1}{2}\sigma_{z}$ and the corresponding projectors of $A_{3,x}$ are  $\{P_{A_{3,x}}^{a}\}$.  We can then have
\begin{align}
\label{pt}
	\frac{\mathbb{I}}{2}=\frac{1}{2}(P_{A_{3,x}}^{+}+P_{A_{3,x}}^{-})
\end{align}

 with $Tr[P_{A_{3,x}}^{+} P_{A_{3,x}}^{-}]=0$. In an ontological model of quantum theory, using convexity property, one can write 
\begin{align}
\label{ptlambda}
	\mu_{P_{x}}(\lambda|\frac{\mathbb{I}}{2})=\frac{1}{2}\left(	\mu_{P_{x}}(\lambda|P_{A_{3,x}}^{+}) +	\mu_{P_{x}}(\lambda|P_{A_{3,x}}^{-})\right)
\end{align}
In a preparation non-contextual model, one assumes $\mu_{P_{x}}(\lambda|\frac{\mathbb{I}}{2})$ is independent of the preparation procedures $\{P_{x}\}\in \mathcal{P}$ where $x=\{1,2,3\}$. As clarified before, we call them the \emph{trivial} preparation non-contextuality assumption. 

The maximally mixed state $\frac{\mathbb{I}}{2}$ can also be prepared by two more non-trivial preparation procedures $P_{4}, P_{5}\in \mathcal{P}$ by measuring suitable POVMs of the following form
\begin{align}
\label{pnt1}
	\frac{\mathbb{I}}{2}=\frac{1}{3}\sum_{x=1}^{3}P_{A_{3,x}}^{+}; \ \ \ \ \frac{\mathbb{I}}{2}=\frac{1}{3}\sum_{x=1}^{3}P_{A_{3,x}}^{-}
		\end{align}
		This is possible only when the functional relations  $A_{3,1} +A_{3,2} +A_{3,3}=0$ is satisfied.
		
Using the convexity property of the $\lambda$ distributions, from Eq.(\ref{pnt1}) one has
\begin{align}
\label{p45}
	&&\mu_{P_{4}}(\lambda|\frac{\mathbb{I}}{2})=\frac{1}{3}\sum_{x=1}^{3}\mu_{P_{4}}(\lambda|P_{A_{3,x}}^{+})\\
	\nonumber
	&&	\mu_{P_{5}}(\lambda|\frac{\mathbb{I}}{2})=\frac{1}{3}\sum_{x=1}^{3}\mu_{P_{5}}(\lambda|P_{A_{3,x}}^{-}) 
\end{align}

which provide the \emph{non-trivial} preparation non-contextuality assumptions. Note that without such non-trivial conditions, logical proof of preparation contextuality for maximally mixed qubit state cannot be revealed \cite{spek05}. 

In a preparation non-contextual model, it is assumed that the $\lambda$ distributions corresponding to the five preparation procedures $\{P_{x}\}$ are the same. Thus, $\mu_{P_{1}}(\lambda|\frac{\mathbb{I}}{2})=\mu_{P_{2}}(\lambda|\frac{\mathbb{I}}{2})=\mu_{P_{3}}(\lambda|\frac{\mathbb{I}}{2})=\mu_{P_{4}}(\lambda|\frac{\mathbb{I}}{2})=\mu_{P_{5}}(\lambda|\frac{\mathbb{I}}{2})\equiv\nu(\lambda|\frac{\mathbb{I}}{2})$. Assume that there exist a $\lambda$ for which $\nu(\lambda|\frac{\mathbb{I}}{2})>0$ and for the same $\lambda$ assume $\mu(\lambda|\{P_{A_{3,x}}^{+}\})>0$. Then by using $\mu(\lambda|P_{A_{3,x}}^{+}) \mu(\lambda|P_{A_{3,x}}^{-})=0$, one finds $\mu_{P_{5}}(\lambda|\frac{\mathbb{I}}{2})=0$ which is in contradiction with the above assignment. A similar contradiction can be found for any assignment of a positive probability distribution for any $\lambda$.

Now, consider that Bob also measures three observables $B_{3,1}, B_{3,2}$ and $ B_{3,3}$. One can then propose a Bell expression is of the form 

\begin{align}
\label{b3}
	\Delta_{3}&= (-A_{3,1} +A_{3,2} +A_{3,3})B_{3,1} + (A_{3,1} -A_{3,2} +A_{3,3})B_{3,2}\\
	\nonumber
	&+  (A_{3,1} +A_{3,2} -A_{3,3}) B_{3,3} 
\end{align}
 
In quantum theory $(\Delta_{3})_{Q}\leq 6$. The maximum value can be achieved if  $B_{3,1}=-A_{3,1}$, $B_{3,2}=-A_{3,2}$ and $B_{3,3}=-A_{3,3}$ are satisfied. 

We now examine the maximum achievable value of $\Delta_{3}$ in local and preparation non-contextual models. In a local model, $(\Delta_{3})_{local}\leq 5$. For the two-party, two-outcome Bell scenario, the violation of locality implies the violation of trivial preparation non-contextuality \cite{pusey}. But, given the observable choices in quantum theory maximizing 	$\Delta_{3}$, Alice's observables  satisfy $A_{3,1} +A_{3,2} +A_{3,3}=0$. In an ontological model of quantum theory, Eq. (\ref{p45}) needs to be satisfied, which provides the non-trivial preparation non-contextuality assumption. If the non-trivial non-contextuality assumption is imposed, we have $(\Delta_{3})_{pnc}\leq 4$. Thus, within the range $4\leq(\Delta_{3})_{Q}< 5$, quantum nonlocality cannot be demonstrated, but a form of non-classicality in terms of non-trivial preparation contextuality can be revealed. 

    We argue that the non-trivial preparation non-contextuality leads us to conclude that Alice cannot steer Bob's state. The argument goes as follows. Each element of the Bob's assemblage $\{\sigma_{\alpha|x}=P_{A_{3,x}}^{a}/6\}$ obtained from Alice's non-trivial preparation  defined by Eq. (\ref{p45}) requires to admit the decomposition given by Eq. (\ref{steer}). Since every $p_{A}(a|x, \lambda)=1$, we found that $\pi(\lambda)=1/6$. This means $\{\sigma_{a|x}\}$ admits the decomposition in Eq. (\ref{steer}) and Alice cannot steer Bob's state. 

By considering the functional relation between Alice's observables, Eq.(\ref{b3}) can be written as $(\Delta_{3})=2\sum_{x=1}^{3} A_{3,x}\otimes B_{3,x}$. Using the preparation non-contextual bound  $(\Delta_{3})_{pnc}\leq 4$, we find the steering inequality is given by
 
 \begin{align}
(\Delta_{3})_{unsteer}=\frac{1}{2}\sum_{x=1}^{3} A_{3,x}\otimes B_{3,x}\leq 1	 
 \end{align}
	which is violated for the choices of observables given above. 

A connection with joint measurability can also be found. Consider Alice measures spin-POVMs defined as $M_{A_{3,x}}^{\pm}=\frac{\mathbb{I}\pm \eta A_{3,x}}{2}$ where $\eta$ is the unsharpness parameter. If Alice performs unsharp measurements of her observables, then the maximum quantum value of the Bell expression given by Eq. (\ref{b3}) is $(\Delta_{3})_{Q}^{max}= 6\eta$. In such a case, nonlocality cannot be demonstrated if $\eta\leq 5/6$, but preparation contextuality may still be revealed. Now, the preparation non-contextuality bound will not be violated if $\eta\leq 2/3$, which is the triple-wise joint measurability condition for trine-spin axes observables.

\subsection{Example of elegant Bell expression}
In this bipartite Bell scenario Alice performs four dichotomic observables $\{A_{3,x}\}$ where $x=\{1,2,3,4\}$ Bob measures the three observables $\{B_{3,y}\}$ where $y=\{1,2,3\}$. One can then write down Gisin's elegant Bell expression is given by

\begin{align}
\label{b43}
	\mathcal{B}_{3}&= (A_{3,1} +A_{3,2} +A_{3,3}-A_{3,4})\otimes B_{3,1} \\
	\nonumber
	&+ (A_{3,1} +A_{3,2} -A_{3,3} +A_{3,4})\otimes B_{3,2} \\
	\nonumber
	&+  (A_{3,1} -A_{3,2} +A_{3,3} +A_{3,4})\otimes B_{3,3} 
\end{align}
Consider Alice's four preparation procedures $\{P_x\} \in \mathcal{P}$ where $x=\{1,2,3,4\}$ corresponding to the dichotomic observables $\{A_{3,x}\}$ produce maximally mixed state is given by $\frac{\mathbb{I}}{2}=\frac{1}{2}(P_{A_{3,x}}^{+}+P_{A_{3,x}}^{-})$. This leads to the \emph{trivial} preparation non-contextuallity assumptions in an ontological model.  

Now, let us consider that in quantum theory, Alice's four preparation procedures $\{P_x\} \in \mathcal{P}$ correspond to four observables $A_{3,1} = (\sigma_x + \sigma_y + \sigma_z )/\sqrt{3}$, $A_{3,2} = (\sigma_x + \sigma_y - \sigma_z )/\sqrt{3}$, $A_{3,3} = (\sigma_x - \sigma_y + \sigma_z) /\sqrt{3}$ and $A_{3,4} = (-\sigma_x + \sigma_y + \sigma_z )/\sqrt{3}$. Note that the projector with $+1$ eigenvalue of $\{A_{3,1}\}$ and projector with $-1$ eigenvalue of the other three observables form SIC-POVM. Then the maximally mixed state $\frac{\mathbb{I}}{2}$ can also be prepared by two non-trivial preparation procedures $P_5$ and $P_6$ are of the following form
\begin{align}
\label{pnt4}
	\frac{\mathbb{I}}{2}=\frac{1}{4}\Big(P_{A_{3,1}}^{+}+\sum_{x=2}^{4}P_{A_{3,x}}^{-}\Big); \ \ \ \ \frac{\mathbb{I}}{2}=\frac{1}{4}\Big(P_{A_{3,1}}^{-}+\sum_{x=2}^{4}P_{A_{3,x}}^{+}\Big)
		\end{align}
		Following the  equivalent representation argument in an ontological model, one has		
\begin{subequations}
\begin{eqnarray}
\label{pnc4logical2}
\mu_{P_{5}}(\lambda|\frac{\mathbb{I}}{2})&=&\frac{1}{4}\left(\mu_{P_{5}}\left(\lambda|P_{A_{3,1}}^{+}\right)+\sum_{x=2}^{4}\mu_{P_{5}}\left(\lambda|P_{A_{3,x}}^{-}\right)\right)\\
\label{pnc4logical3}
\mu_{P_{6}}(\lambda|\frac{\mathbb{I}}{2})&=&\frac{1}{4}\left(\mu_{P_{6}}\left(\lambda|P_{A_{3,1}}^{-}\right)+\sum_{i=2}^{4}\mu_{P_{6}}\left(\lambda|P_{A_{3,x}}^{+}\right)\right)
\end{eqnarray}
\end{subequations}
which provides \emph{non-trivial} preparation noncontextuality assumption. Note that Eq. (\ref{pnt4}) is satisfied only if $A_{3,1} =A_{3,2} +A_{3,3}+A_{3,4}$.
		
 If ontological model is preparation non-contextual for mixed states, then $\mu_{P_{1}}(\lambda|\frac{\mathbb{I}}{2})=\mu_{P_{2}}(\lambda|\frac{\mathbb{I}}{2})=\mu_{P_{3}}(\lambda|\frac{\mathbb{I}}{2})=\mu_{P_{4}}(\lambda|\frac{\mathbb{I}}{2})=\mu_{P_{5}}(\lambda|\frac{\mathbb{I}}{2})=\mu_{P_{6}}(\lambda|\frac{\mathbb{I}}{2})=\nu(\lambda|\frac{\mathbb{I}}{2})$. However, no logical proof can be demonstrated in this case.

If $B_{3,1}=\sigma_x$, $B_{3,2}=\sigma_y$ and $B_{3,3}=\sigma_z$, the optimal quantum value $(\mathcal{B}_{3})_{Q}^{max}= 4\sqrt{3}$ \cite{acin,ghorai} can be achieved for the maximally entangled state $|\psi\rangle =\frac{1}{\sqrt{2}}\left(|00\rangle-|11\rangle\right)$.

In a local model $(\mathcal{B}_{3})_{local}\leq 6$. But if non-trivial preparation non-contextuality  corresponding to Eq. (\ref{pnt4}) is assumed a functional relation $A_{3,1} =A_{3,2} +A_{3,3}+A_{3,4}$ needs to be satisfied. By imposing the above condition, the local bound reduces to the non-trivial preparation non-contextual bound is given by $(\mathcal{B}_{3})_{pnc}\leq 4$. Following the similar argument given in \ref{pncsteer} and at the end of \ref{sstrine}, we can argue that the non-trivial preparation non-contextuality assumption leads us to conclude the unsteerability of Bob assemblage by Alice. In other words, the ranges of values of elegant Bell expression $4\leq (\mathcal{B}_{3})_{Q}< 6$ provide the non-classicality in the form of quantum steering. 

The elegant Bell inequality for the local model can be transformed into the well-known linear steering inequality for three measurements as follows. Noting the choices of observables of Alice and Bob for maximizing the elegant Bell expression in quantum theory and providing the non-trivial preparation non-contextuality assumptions, we can write $A_{3,1} +A_{3,2} +A_{3,3}-A_{3,4} =(4/\sqrt{3}) B_{3,1}$, $A_{3,1} +A_{3,2} -A_{3,3}+A_{3,4}=(4/\sqrt{3}) B_{3,2}$ and $A_{3,1} -A_{3,2} +A_{3,3}+A_{3,4} =(4/\sqrt{3}) B_{3,3}$. Putting them in Eq. (\ref{b43}), we have  
\begin{align}
\label{bb}
	\mathcal{B}_{3}=\frac{4}{\sqrt{3}}\sum_{y=1}^{3}  \langle B_{3,y}\otimes B_{3,y}\rangle
\end{align}
Since $(\mathcal{B}_{3})_{pnc}\leq 4$, from Eq. (\ref{bb}) we have ,

 \begin{align}
(	\mathcal{B}_{3})_{unsteer}=\frac{1}{\sqrt{3}}\sum_{y=1}^{3} \langle B_{3,y}\otimes B_{3,y}\rangle \leq 1
\end{align}
which is the well-known linear steering inequality for three measurement scenario \cite{jones} 

If Bob measures unsharp spin-POVMs $E_{b|y}=(\mathbb{I}+ b\eta B_{3,y})/2$, we have $(\mathcal{B}_{3})_{Q}=4\sqrt{3}\eta$. For the violation of locality one requires $\eta>\frac{\sqrt{3}}{2}\approx 0.86$. On the other hand, no violation of preparation non-contextuality can be obtained for $\eta\leq \frac{1}{\sqrt{3}}$ which is the triple-wise joint measurability condition for the observables corresponding to three orthogonal axes.   
 
\section{Summary and Discussion}
 For a bipartite binary measurement Bell scenario, a link between quantum nonlocality and trivial preparation contextuality has already been demonstrated \cite{barrett,pusey,tava19}. This work provided a direct connection between quantum steering and non-trivial preparation contextuality. The trivial preparation non-contextuality assumption arises due to the projective measurement of Alice. On the other hand, the non-trivial ones appear when Alice performs the measurements of suitable POVMs originating from a suitably chosen set of functional relations between Alice's observables. Further, we have shown that certain local Bell inequalities can be cast into the non-trivial preparation non-contextuality inequalities. The latter inequalities can also be considered linear steering inequalities. Thus, the same Bell expression can demonstrate the nonlocality and quantum steering. 

We considered two local Bell inequalities to demonstrate our results, which provide maximum violation for the qubit system. We first showed that the local Bell's inequality involving three dichotomic observables in each party could be converted into a linear steering (or non-trivial preparation non-contextual) inequality by using the non-trivial functional relation between Alice's observables. We then showed how Gisin's elegant Bell inequality \cite{gisin} can also be converted into a known linear steering inequality for three-measurement. 

 Note that the equivalence representation of the parity-oblivious condition in quantum theory provides an ontological model non-trivial preparation non-contextuality assumption. We argued that due to the existence of such non-trivial constraints on Alice's choices of observables, the assemblage created by Alice to Bob is unsteerable. In such a case, the aforementioned family of local Bell inequalities can be cast into two types of new linear steering inequalities. In other words, the steering inequalities derived here can also be considered as the non-trivial preparation non-contextuality inequalities. Thus we provide a scheme for testing the quantum steering and nonlocality from the same family of Bell expressions. 
 
 Recently, one sided device-independent tasks have found to be fundamental resources in the field of quantum information theory in the steering context \cite{sch,reid,wise1,jones,cava}. Further, quantum contextuality also provides advantage in information processing tasks such as communication games \cite{spek09,hameedi,ghorai}
and in quantum computation \cite{rau,how}. One such advantage in communication games can be illustrated in parity-oblivious random-access-code \cite{spek09}. Since, parity-oblivious condition in quantum theory has a number of applications which showcase advantage in tasks over classical theory, its ontological model satisfying non-trivial preparation non-contextuality assumption has implications in the field of quantum communication and quantum computation. Thus, our work establishes the link between steering and non-trivial preparation non-contextuality, thereby paves the path for exploring range of new applications in the field of quantum information theory.  
  
\section*{Data availability} Data sharing is not applicable to this article as no datasets were generated or analyzed during the current study.

\section*{Author contribution statement}
AKP conceived the idea and supervised the work. Both authors contributed to the calculations, and preparation of the manuscript.

\section*{Acknowledgments}
PR acknowledges the support from the research grant DST/ICPS/QuEST/2019/4. AKP acknowledges the support from the research grant MTR/2021/000908.


\begin{thebibliography}{99}
\bibitem{bell64}J. S. Bell, Physics 1 195 (1964).
\bibitem{bell66}J. S. Bell, Rev. Mod. Phys. 38, 447 (1966).
\bibitem{ks}S. Kochen and E. P. Specker, J. Math. Mech. 17, 59 (1967).
\bibitem{marco05} M. Genovese, Physics Reports 413, 6, 319-396 (2005).
\bibitem{Brunner2014} N. Brunner \emph{et al.}, Rev. Mod. Phys. 86, 419 (2014).

\bibitem{marco19} M. Genovese and M. Gramegna, Appl. Sci. 9 (24), 540 (2019).
\bibitem{ekert} A. K. Ekert,  Phys. Rev. Lett. 67, 661 (1991).
\bibitem{acin} A. Acin \emph{et al.},   Phys. Rev. Lett. 98, 230501 (2007).
\bibitem{bar13} J. Barrett, R. Colbeck and A. Kent, Phys. Rev. Lett. 110, 010503 (2013).
\bibitem{zap22} V. Zapatero \emph{et al.},	arXiv:2208.12842 (2022).
\bibitem{pir10} S. Pironio \emph{et al.},  Nature 464, 1021–1024 (2010).
\bibitem{acin12}A. Acin, S. Massar and S. Pironio,  Phys. Rev. Lett. 108, 100402 (2012).
\bibitem{col12}R. Colbeck and R. Renner, Nature Phys. 8, 450–453 (2012).
\bibitem{pathak22} V. Mannalath, S. Mishra and A. Pathak,  	arXiv:2203.00261 (2022).
\bibitem{supic}I. Supic and J. Bowles, Quantum 4, 337 (2020).
\bibitem{dim} N. Brunner \emph{et al.}, T Phys. Rev. Lett. 100, 210503 (2008).
\bibitem{mahato}A. K. Pan and S. S. Mahato, Phys. Rev. A 102, 052221 (2020). 
 \bibitem{spek09}R. W. Spekkens \emph{et al.}, Phys. Rev. Lett. 102, 010401 (2009).	
 \bibitem{hameedi}A. Hameedi, A. Tavakoli, B. Marques and M. Bourennane,  Phys. Rev. Lett. 119, 220402 (2017).
\bibitem{ghorai}S. Ghorai and A. K. Pan,  Phys. Rev A, 98, 032110 (2018). 
 \bibitem{um13} M. Um \emph{et al.}, Sci. Rep. 3, 1627 (2013).
 \bibitem{um20} M. Um \emph{et al.}, Phys. Rev. Applied 13, 034077 (2020).
 \bibitem{pan21d} A. K. Pan, Eur. Phys. J. D 75, 98 (2021).
 \bibitem{rau} R. Raussendorf, Phys. Rev. A 88, 022322 (2013).
\bibitem{how} M. Howard, J. Wallman, V. Veitch and J. Emerson,  Nature, 510, 351 (2014).
\bibitem{mermin} N. D. Mermin,  Rev. Mod. Phys. 65, 803 (1993).
\bibitem{ker}M. Kernaghan,  J. Phys. A 27, L829 (1994).
\bibitem{cabello} A. Cabello, J. M. Estebaranz, and G. Garcia-Alcaine,  Phys. Lett. A 212, 183 (1996).
\bibitem{cabello08} A. Cabello,  Phys. Rev. Lett.  101, 210401 (2008).
\bibitem{kly} A. A. Klyachko, M. A. Can, S. Binicioglu and A. S. Shumovsky,  Phys. Rev. Lett. 101, 020403 (2008).
\bibitem{pan10} A. K. Pan,   EPL 90, 40002 (2010).
\bibitem{yu} S. Yu and C.H. Oh,  Phys. Rev. Lett. 108, 030402 (2012).
\bibitem{pan21s} A. K. Pan, Sci. Rep. 9, 17631 (2019).
\bibitem{budroni21} C. Budroni \emph{et al.},	arXiv:2102.13036 (2021).
\bibitem{spek05}R. W. Spekkens,  Phys. Rev. A 71, 052108 (2005).
\bibitem{hari} N. Harrigan and R. Spekkens,  Found. Phys. 40, 125(2010).
\bibitem{spek14} R. Spekkens,  Found. Phys. 44, 1125 (2014).
\bibitem{kumari} A. Kumari and A. K. Pan, Phys. Rev. A 100, 062130 (2019).
\bibitem{epr} A. Einstein, B. Podolsky, and N. Rosen, Phys. Rev.
47, 777 (1935).
\bibitem{sch} E. Schrodinger, Proc. Cambridge Philos. Soc. 31, 555 (1935).
\bibitem{reid}M. D. Reid, Phys. Rev. A 40, 913 (1989).
\bibitem{wise1} H. M. Wiseman, S. J. Jones, and A. C. Doherty, Phys. Rev. Lett. 98, 140402 (2007).
\bibitem{jones}S. J. Jones, H. M. Wiseman, and A. C. Doherty, Phys. Rev. A. 76, 052116 (2007).
\bibitem{cava}E. G. Cavalcanti, S. J. Jones, H. M. Wiseman, and M. D. Reid, Phys. Rev. A 80, 032112 (2009).
\bibitem{quin}M. T. Quintino, T. Vertesi, and N. Brunner, Phys. Rev. Lett. 113, 160402 (2014).
\bibitem{uolaprl} R. Uola, T. Moroder, and O. Guhne, Phys. Rev. Lett. 113, 160403 (2014).
\bibitem{pati} J-L. Chen, H-Y. Su, Z-P. Xu and  A.K. Pati, Sci. Rep. 6, 32075 (2016).
\bibitem{dd1}D. Das, S. Sasmal, and S. Roy,
Phys. Rev. A 99, 052109 (2019).
\bibitem{dd2}D. Das, S. Sasmal, and A. Roy, Quantum Inf. Processing 18, 315 (2019).
\bibitem{uolarev} R. Uola, Ana C. S. Costa, H. Chau Nguyen, and O. Guhne, Rev. Mod. Phys. 92, 015001 (2020).
\bibitem{barrett} J. Barrett (unpublished, private communication). 
\bibitem{pusey}M. Pusey,  Phys. Rev. A 98, 022112 (2018).
\bibitem{tava19} A. Tavakoli and R. Uola, Phys. Rev. Research 2, 013011 (2020).
\bibitem{gisin} N.Gisin, arXiv:quant-ph/0702021.
\bibitem{saunders}D. J. Saunders, S. J. Jones, H. M. Wiseman, and G. J. Pryde, Nature Phys. 6, 845–849 (2010).


\end{thebibliography}
\end{document}